\documentclass[a4paper, 11pt]{amsart}
\usepackage{graphicx}
\usepackage{bm}
\usepackage{amsmath}
\usepackage{amsfonts}
\usepackage{amssymb}

\begin{document}

\title{Informationally complete characters for quark and lepton mixings}

\author{Michel Planat$\dag$, Raymond Aschheim$\ddag$,\\ Marcelo M. Amaral$\ddag$ and Klee Irwin$\ddag$}

\address{$\dag$ Universit\'e de Bourgogne/Franche-Comt\'e, Institut FEMTO-ST CNRS UMR 6174, 15 B Avenue des Montboucons, F-25044 Besan\c con, France.}
\email{michel.planat@femto-st.fr}

\address{$\ddag$ Quantum Gravity Research, Los Angeles, CA 90290, USA}
\email{raymond@QuantumGravityResearch.org}
\email{Klee@quantumgravityresearch.org}
\email{Marcelo@quantumgravityresearch.org}

\begin{abstract}

A popular account of the mixing patterns for the three generations of quarks and leptons is through the characters  $\kappa$ of a finite group $G$. Here we introduce a $d$-dimensional Hilbert space with $d=cc(G)$, the number of conjugacy classes of $G$. Groups under consideration should follow two rules, 
(a) the character table contains both two- and three-dimensional representations with at least one of them faithful and (b) there are minimal informationally complete measurements under the action  of a $d$-dimensional Pauli group over the characters of these representations. Groups with small $d$ that satisfy these rules coincide in a large part with viable ones derived so far for reproducing simultaneously the CKM (quark) and PNMS (lepton) mixing matrices. Groups leading to physical $CP$ violation are singled out.

\end{abstract}

\maketitle

\vspace*{-.5cm}
\footnotesize {~~~~~~~~~~~~~~~~~~~~~~PACS: 03.67.-a, 12.15.Ff, 12.15.Hh, 03.65.Fd, 98.80.Cq }


\footnotesize {Keywords: Informationally complete characters, quark and lepton mixings, CP violation, Pauli groups}

\normalsize

\section{Introduction}

\begin{figure}[ht]
\includegraphics[width=8cm]{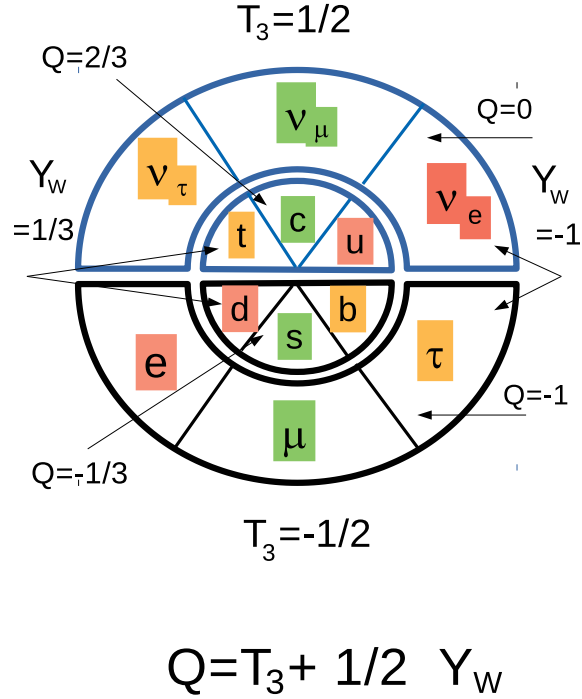}
\caption{An angular picture of the three generations of quarks and leptons. The blue and black pancakes have isospin is $1/2$ and $-1/2$, respectively. The inside and outside crowns have weak hypercharges $\frac{1}{3}$ and $-1$, respectively.
}
\label{partic}
 \end{figure}

In the standard model of elementary particles and according to the current experiments there exist three generations of matter but we do not know why. The matter particles are fermions of spin $1/2$ and comprise the quarks (responsible for the strong interactions) and leptons (responsible for the electroweak interactions as shown in Table \ref{tab0} and Fig. \ref{partic}.

\begin{table}[h]
\begin{center}
\begin{tabular}{|c||l|c|l||c|l|c||}
\hline 
\hline
matter & type 1 & type 2 & type 3 &  Q & $T_3$ &$Y_W$\\
\hline
(1) quarks   & u & c & t & 2/3  &  1/2 &1/3\\
         & d & s & b & -1/3 & -1/2 & .\\
\hline
(2) leptons  & e & $\mu$ & $\tau$ & -1 & -1/2 & -1\\
         & $\nu_e$ & $\nu_{\mu}$ & $\nu_{\tau}$ & 0 & 1/2 & .\\
\hline
\hline
\end{tabular}
\caption{(1) The three generations of up-type quarks (up, charm and top) and of down-type quarks (down, strange and bottom),
(2) The three generations of leptons (electron, muon and tau) and their partner neutrinos. The symbols $Q$, $T_3$ and $Y_W$ are for the charge, the isospin and the weak hypercharge, respectively. They satisfy the equation $Q=T_3+\frac{1}{2}Y_W$.} 
\label{tab0}
\end{center}
\end{table}

In order to explain the CP-violation (the non-invariance of interactions under the combined action of charged-conjugation (C) and parity (P) transformations) in quarks, Kobayashi and Maskawa introduced the so-called  Cabibbo-Kobayashi-Maskawa unitary matrix (or CKM matrix) that describes the probability of transition from one quark $i$ to another $j$. These transitions are proportional to $|V_{ij}|^2$ where the $V_{ij}$'s are entries in the CKM matrix \cite{Ramond2011, Holthausen2013}

\begin{equation*}
U_{CKM}=
\begin{pmatrix}   V_{ud} & V_{us} & V_{ub} \\
                  V_{cd} & V_{cs} & V_{cb} \\
                  V_{td} & V_{ts} & V_{tb} \\
									\end{pmatrix}
		\mbox{with}~ |U_{CKM}|~	\approx 		
							\begin{pmatrix}   0.974 & 0.225 & 0.004 \\
                  0.225 & 0.973 & 0.041 \\
                  0.009 & 0.040 & 0.999 \\
									\end{pmatrix}.
\end{equation*}

There is a standard parametrization of the CKM matrix with three Euler angles $\theta_{12}$ (the Cabbibo angle), $\theta_{23}$, $\theta_{12}$	and the CP-violating phase $\delta_{CP}$. 
Taking $s_{ij}=\sin(\theta_{ij})$ and $c_{ij}=\cos(\theta_{ij})$, the CKM matrix reads		

\begin{equation*}								
\begin{pmatrix} 1 & 0 & 0\\
                  0 & c_{23} & s_{23} \\
                  0 & -s_{23} & c_{23} \\ \end{pmatrix}
									\begin{pmatrix} c_{13} & 0 & s_{13}e^{-i\delta_{CP}}\\
                  0 & 1 & 0 \\
                  s_{13}e^{-i\delta_{CP}} & 0 & c_{13} \\ \end{pmatrix}
									\begin{pmatrix} c_{12} & s_{12} & 0\\
                  -s_{12} & c_{12} & 0 \\
                  0 & 0 & 1 \\
									\end{pmatrix}.								
\end{equation*}

Similarly, the charged leptons $e$, $\mu$ and $\tau$ partner with three generations of flavors of neutrinos $\nu_e$, $\nu_\mu$ and $\nu_\tau$ in the charged-current weak interaction. Neutrino's mass  $m_i$ can be deduced with probability $|U_{\alpha i}|^2$ where the $U_{\alpha i}$'s are the amplitudes of mass eigenstates $i$ in flavor $\alpha$. The so-called Pontecorvo–Maki–Nakagawa–Sakata unitary matrix (or PMNS matrix) is as follows \cite{NeutrinoFit}

\begin{equation*}
 U_{PMNS}=	
\begin{pmatrix}
 U_{e 1 }   & U_{e 2}    & U_{e 3} \\
 U_{\mu 1}  & U_{\mu 2}  & U_{\mu 3} \\
 U_{\tau 1} & U_{\tau 2} & U_{\tau 3} \\ 
\end{pmatrix}
 \end{equation*}		

\begin{equation*}
\mbox{with}~ |U_{PMNS}|\approx
\begin{pmatrix}
 0.799 \rightarrow 0.844   & 0.516 \rightarrow 0.582     & 0.141 \rightarrow 0.156 \\
 0.242\rightarrow 0.494  & 0.467\rightarrow 0.678   & 0.639\rightarrow 0.774  \\
 0.284\rightarrow 0.521  & 0.490\rightarrow 0.695  & 0.615 \rightarrow 0.754  \\ 
\end{pmatrix},
\end{equation*}			

\noindent
where the entries in the matrix mean the range of values	allowed by the present day experiments.
																	
As for the CKM matrix the three mixing angles are denoted	$\theta_{12}$, $\theta_{23}$, $\theta_{12}$	and the CP-violating phase is called $\delta_{CP}$.

The current experimental values of angles for reproducing entries in the CKM and PMNS matrices are in Table \ref{tabangles}.

\begin{table}[h]
\begin{center}
\begin{tabular}{|c|l|c|l|c|l|}
\hline 
\hline
angles (in degrees)& $\theta_{12}$ & $\theta_{13}$ & $\theta_{23}$ &  $\delta_{CP}$ \\
\hline
quark mixings & 13.04 & 0.201 & 2.38 & 71\\
\hline
lepton mixings
 & 33.62 & 8.54 & 47.2 & -90 \\
\hline
\hline
\end{tabular}
\caption{Experimental values of the angles in degrees for mixing patterns of quarks (in the CKM matrix) and leptons (in the PMNS matrix). } 
\label{tabangles}
\end{center}
\end{table}

In the last twenty years or so, a paradigm has emerged that it may exist an underlying discrete symmetry jointly explaining the mixing patterns of quarks and leptons \cite{Frampton2007,Xing2020}. 
This assumption follows from the fact that the CKM matrix is found closed to the identity matrix and the entries in the PMNS matrix are found to be of order one except for the almost vanishing $U_{e3}$. A puzzling difference between quark and lepton mixing lies in the fact that there is much more neutrino mixing than mixing between the quark flavors. Up and down quark matrices are only slightly misaligned while there exists a strong misalignment of charged leptons with respect to neutrino mass matrices. A valid model should account for these features.
 
The standard model essentially consists of two continuous symmetries, the electroweak symmetry $SU(2) \times U(1)$ (that unifies the electromagnetic and weak interactions) and the quantum chromodynamics symmetry $SU(3)$ (that corresponds to strong interactions). There are several puzzles not explained within the standard model including the flavor mixing patterns, the fermion masses, and the CP violations in the quark and lepton sectors. There are astonishing numerical coincidences such as the Koide formula for fermion masses \cite{Sheppeard2017}, the quark-lepton complementarity relations $\theta_{12}^{\mbox{quark}}+\theta_{12}^{\mbox{lepton}} \approx \pi/4$, $\theta_{23}^{\mbox{quark}} \pm \theta_{23}^{\mbox{lepton}}\approx \pi/4$ \cite{Minakata2004} and efficient first order models such as the tribimaximal model \cite{Parattu2011}-\cite{ChenChuliaDing2018} and the \lq Golden ratio' model \cite{Kajiyama2007,QGRIAAF}. For instance, tribimaximal mixing gives values of angles as
$\theta_{12}^{\mbox{lepton}}=\sin^{-1}(\frac{1}{\sqrt{3}})\approx 35.3^{\circ}$, $\theta_{23}^{\mbox{lepton}}=45^{\circ}$, $\theta_{13}^{\mbox{lepton}}=0$ and $\delta_{CP}=0$, compatible with earlier data. Such a model could be made more realistic by taking two CP-phases instead of one \cite{ChenChuliaDing2018}.

Currently many discrete models of quark-lepton mixing patterns are based on the representations of finite groups that are both subgroups of $U(2)$ and $U(3)$ \cite{Jurciukonis2017}-\cite{Hagedorn2014}. In the same spirit, we add to this body of knowledge by selecting valid subgroups of unitary groups from a criterion borrowed to the theory of generalized quantum measurements. 

One needs a quantum state (called a fiducial state) and one also requires that such a state is informationally complete under the action of a $d$-dimensional Pauli group $\mathcal{P}_d$. When such a state is not an eigenstate of a $d$-dimensional Pauli group it allows to perform universal quantum computation \cite{PlanatGedik}-\cite{MPGQR1}. In the latter papers, valid states belong to the eigenstates of mutually commuting permutation matrices in a permutation group derived from the coset classes of a free group with relations. From now, the fiducial state will have to be selected from the characters $\kappa$ of a finite group $G$ with the number of conjugacy classes $d=cc(G)$ defining the Hilbert space dimension. Groups under consideration should obey two rules (a) the character table of $G$ contains both two- and three-dimensional representations with at least one of them faithful and (b) there are minimal informationally complete measurements under the action  of a $d$-dimensional Pauli group over the characters of  these representations. The first criterion is inspired by the current understanding of quark and lepton mixings (and the standard model) and the second one by the theory of magic states in quantum computing \cite{PlanatGedik}. Since matter particles are spin $1/2$ fermions it is entirely consistent to see them under the prism of quantum measurements.

In the rest of this introduction we recall what we mean by a minimal informationally complete quantum measurement (or MIC). In Section \ref{MICQuarkLepton} we apply criteria (a) and (b) to groups with small $cc \le 36$ where we can perform the calculations. Then we extrapolate to some other groups with $cc>36$. Most groups found from this procedure fit the current literature as being viable for reproducing lepton and quark mixing patterns. In Section \ref{CPseries}, we examine the distinction between generalized CP symmetry and CP violation and apply it to our list of viable groups.

\subsection*{Minimal informationally complete quantum measurements}

Let $\mathcal{H}_d$ be a $d$-dimensional complex Hilbert space and $\{E_1,\ldots,E_m\}$ be a collection of positive semi-definite operators (POVM) that sum to the identity. Taking the unkwown quantum state as a rank one projector $\rho=\left|\psi\right\rangle \left\langle \psi \right|$ (with $\rho^2=\rho$ and $\mbox{tr}(\rho)=1$), the $i$-th outcome is obtained with a probability given by the Born rule $p(i)=\mbox{tr}(\rho E_i)$. 
A minimal and informationally complete POVM (or MIC) requires $d^2$ one-dimensional projectors $\Pi_i=\left|\psi_i\right\rangle \left\langle \psi_i \right|$, with $\Pi_i=d E_i$, such that the rank of the Gram matrix with elements $\mbox{tr}(\Pi_i\Pi_j)$, is precisely $d^2$.

With a MIC, the complete recovery of a state $\rho$ is possible at a minimal cost from the probabilities $p(i)$. In the best case, the MIC is symmetric and called a SIC with a further relation $\left |\left\langle \psi_i|\psi_j \right \rangle \right |^2=\mbox{tr}(\Pi_i\Pi_j)=\frac{d\delta_{ij}+1}{d+1}$ so that the density matrix $\rho$ can be made explicit \cite{Fuchs2004,Fuchs2020}.

In our earlier references \cite{PlanatGedik,PlanatModular}, a large collection of  MICs are derived. They correspond to Hermitian angles $\left |\left\langle \psi_i|\psi_j \right \rangle \right |_{i \ne j} \in A=\{a_1,\ldots,a_l\}$ belonging a discrete set of values of small cardinality $l$.
They arise from the action of a Pauli group $\mathcal{P}_d$ \cite{Pauli2011} on an appropriate magic state pertaining to the coset structure of subgroups of index $d$ of a free group with relations.

Here, an entirely new class of MICs in the Hilbert space $\mathcal{H}_d$, relevant for the lepton and quark mixing patterns,  is obtained by taking fiducial/magic states as characters of a finite group $G$ possessing $d$ conjugacy classes and using the action of a Pauli group $\mathcal{P}_d$ on them.


\section{Informationally complete characters for quark/lepton mixing matrices}
\label{MICQuarkLepton}

The standard classification of small groups is from their cardinality. Finite groups relevant to quark and lepton mixings are listed accordingly \cite{Parattu2011, Jurciukonis2017,YaoDingadd}. We depart from this habit by classifying the small groups $G$ of interest versus the number $d=cc(G)$ of their conjugacy classes. This motivation is due to the application of criterion (b) where we need to check whether the action of a Pauli group in the $d$-dimensional Hilbert space $\mathcal{H}_d$ results in a minimal informationally complete POVM (or MIC).

\begin{table}[h]
\begin{center}
\begin{tabular}{|c|l|c|l|c|l|}
\hline 
\hline
Group & Name or signature & $cc$ &  Graph & Ref  \\
\hline
SmallGroup(24,12) & $S_4$, $\Delta(6 \times 2^2)$  & 5 & $K_4$ & \cite{Jurciukonis2017} \\  
SmallGroup(120,5) & 2I, SL(2, 5) & 9 & $K_5^3$ & \cite{Hashimoto2011},$\dag$, $\ddag$\\
SmallGroup(150,5) & $\Delta(6 \times 5^2)$ &13 & $ K_5^3$& \cite{Holthausen2013,Jurciukonis2017,King2013}\\
SmallGroup(72,42) & $\mathbb{Z}_4 \times S_4$ & 15 & $K_3^4$ & \cite{Parattu2011}\\
SmallGroup(216,95)& $\Delta(6 \times 6^2)$ &19 & $K_6^3$ & \cite{Jurciukonis2017}\\
SmallGroup(294,7) & $\Delta(6 \times 7^2)$ &20 & ? & \cite{LiDing2015}\\
SmallGroup(72,3) & $Q_8 \rtimes \mathbb{Z}_9$ & 21 & $K_2^3$ &\cite{Parattu2011}\\
SmallGroup(162,12) & $\mathbb{Z}_3^2 \rtimes(\mathbb{Z}_3^2 \rtimes \mathbb{Z}_2)$ & 22 & $K_9^3$ &\cite{Holthausen2013,Jurciukonis2017,YaoDing2015}\\
SmallGroup(162,14) & ., $D_{9,3}^{(1)}$ & . & . & \cite{Holthausen2013,Jurciukonis2017,LiYaoDing2016} \\
SmallGroup(384,568) & $\Delta(6 \times 8^2)$ &24 & ? & \cite{LiDing2015}\\
SmallGroup(648,532) & $\Sigma(216 \times 3)$, $\mathbb{Z}_3 \rtimes(\mathbb{Z}_3 \rtimes SL(2,3))$& 24& ?&\cite{Jurciukonis2017,Hagedorn2014}\\
SmallGroup(648,533) & Q(648) , .& 24& ?&\cite{Jurciukonis2017,KingLudl2016}\\
SmallGroup(120,37) &  $\mathbb{Z}_5 \times S_4$& 25 & $K_5^4$ & $\dag$ \\    
SmallGroup(360,51) & $\mathbb{Z}_3 \times SL(2,5)$& 27 & $K_{12}^6$& $\dag$\\ 
SmallGroup(162,44) &    $\mathbb{Z}_3^2 \rtimes(\mathbb{Z}_3^2 \rtimes \mathbb{Z}_2)$ & 30 & $K_9^3$ & \cite{Jurciukonis2017}\\   
SmallGroup(600,179) & $\Delta(6 \times 10^2)$ & 33 & $K_{10}^3$ &\cite{Holthausen2013,Jurciukonis2017,King2013} \\  
SmallGroup(168,45) &  $\mathbb{Z}_7 \times S_4$  & 35 & $K_7^4$ & $\dag$ \\  
SmallGroup(480,221) & $\mathbb{Z}_8. A_5$, $SL(2,5).\mathbb{Z}_4$  &36 & $K_8^6$ & $\ddag$ \\                        
\hline
\hline
\end{tabular}
\caption{List of the $16+2$ groups with number of conjugacy classes $cc \le 36$ that satisfy rules (a) and (b). Groups $(294,7)$ and $(384,568)$ need two CP phases to become viable models as mentioned in Section \ref{CPseries}. The smallest permutation representation on $k\times l$ letters stabilizes the $n$-partite graph $K_k^l$ given at the fourth column. The groups $\Delta(6 \times n^2)$ is isomorphic to $\mathbb{Z}_n^2 \rtimes S_3$. A reference is given at the last column if a viable model for quark or/and lepton mixings can be obtained. The extra cases with reference $\dag$ and $\ddag$ can be found in \cite{Chaber2018add} and \cite{YaoDingadd}, respectively.} 
\label{table1}
\end{center}
\end{table}

A list of finite groups $G$ according to the number of their conjugacy classes (complete only up to $d \le 12$) is in Ref. \cite{Lopez1985}.
 It can also be easily recovered with a simple code in Magma or Gap. For our application to quark and lepton mixings, we need much higher $d$. In practice, we used existing tables of subgroups of $U(3)$ (of cardinality up to $2000$ in \cite{Parattu2011,Jurciukonis2017,YaoDingadd} and up to $1025$ in \cite{Chaber2018} to select our group candidates). 

Tables \ref{table1} is the list of $16+2$  small groups with $cc\le 36$ found to satisfy the two rules (a) the character table of $G$ contains both two- and three-dimensional representations with at least one of them faithful and (b) the quantum measurement is informationally complete under a $d$-dimensional Pauli group.

The $16$ groups lead to good models for the absolute values of entries in the CKM and PMNS matrices except for the ones that have the factor $SL(2,5)$ in their signature. The two extra groups  $(294,7)=\Delta(6 \times 7^2)$ and $(384,568)=\Delta(6 \times 8^2)$ arise when one takes into account the generalized CP symmetry as  in Section \ref{CPseries}.

Details are in Table \ref{tablefaithshort} for the first three groups and the group $(294,7)$. Full results are in Table \ref{tablefaith} and \ref{tablefaith2} of the Appendix.

\begin{table}[h]
\begin{center}
\begin{tabular}{||c|l||c|l|c|l|c|l|c|l|c|l|c|l|c||}
\hline 
\hline
Group & d&  &  &  & & & & & & & & & & \\
\hline
(24,12) & 5& 1 & 1 & 2 & {\bf 3} & {\bf 3} & & & & & & && \\
$5$-dit & . & 5 & 21 & $d^2$ & $d^2$ & $d^2$ & & & & & & &&\\        
\hline
(120,5) & 9& 1 & {\bf 2} & {\bf 2} & 3 & 3 & 4 & {\bf 4}& 5& {\bf 6}& & & &\\
$9$-dit & . & 9 & $d^2$ & $d^2$ & $d^2$ & $d^2$ & $d^2$&$d^2$& 79& $d^2$& & & &  \\  
2QT & . & 9 & $d^2$ & $d^2$ & $d^2$ & $d^2$ & $d^2$&$d^2$& $d^2$& $d^2$& & & &  \\    
\hline
(150,5) & 13& 1  &1  &2  &{\bf 3} & {\bf 3}& {\bf 3}& {\bf 3}& {\bf 3}& {\bf 3}& {\bf 3}& {\bf 3}&{\bf 6} & {\bf 6} \\
13-dit & .& 13  &157  & $d^2$  & $d^2$ &  $d^2$&  $d^2$&  $d^2$&  $d^2$&  $d^2$&  $d^2$&  $d^2$& $d^2$ &  $d^2$ \\
\hline
\hline
(294,7) & 20& 1 & 1 & 2 & {\bf 3}&{\bf 3} &{\bf 3} &{\bf 3}&{\bf 3} &{\bf 3}&{\bf 3} &{\bf 3} & {\bf 3}& {\bf 3}\\
20-dit & .& 20 &349  & 388 & $d^2$ &$d^2$ &$d^2$ &$d^2$ &$d^2$ &$d^2$ &$d^2$ & $d^2$& $d^2$&$d^2$ \\
. & .& {\bf 6} & {\bf 6} &{\bf 6} &{\bf 6} &{\bf 6} & & & & & & & & \\
. & .& 390& 390 &390 &398 &398 & & & & & & & & \\
\hline
\hline
\end{tabular}
\caption{The first three small groups considered in our Table \ref{table1} and group $(294,7)$ added in Section \ref{CPseries}. For each group and each character the table provides the dimension of the representation and the rank of the Gram matrix obtained under the action of the corresponding Pauli group. Bold characters are for faithful representations. According to our demand, each selected group has both $2$- and $3$-dimensional characters (with at least one of them faithful) that are fiducial states for an informationally complete POVM (or MIC) with the  rank of Gram matrix equal to $d^2$. The Pauli group performing this action is a $d$-dit or a $2$-qutrit (2QT) for the group $(120,5)=SL(2,5)=2I$.} 
\label{tablefaithshort}
\end{center}
\end{table}
\normalsize

Table \ref{table2} is an extrapolation to groups with higher $cc$ where criterion (a) is satisfied but where (b) could not be checked. Most groups in the two tables have been found to be viable models and several of them belong to known sequences.

In tables \ref{table1} and \ref{table2}, the first column is the standard small group identifier in which the first entry is the order of the group (as in \cite{Jurciukonis2017}). At the second column, one finds a signature in terms of a direct product (with the symbol $\times$), a semidirect product (with the symbol $\rtimes$), a dot product (with the symbol $.$) or a member of a sequence of groups such as the $\Delta(6 \times n^2)$ sequence found to contain many viable groups for quark and lepton mixings. 
The third column gives the number of irreducible characters/conjugacy classes. Another information is about the geometry of the group. To get it, one first selects the smallest permutation representation on $k\times l$ letters of $G$. Then one looks at the two-point stabilizer subgroup $G_s$ of smallest cardinality in the selected group $G$. The incidence matrix of such a subgroup turns out to be the $l$-partite graph $K_k^l$ that one can identify from the graph spectrum. Such a method is already used in our previous papers about magic state type quantum computing \cite{PlanatGedik}-\cite{MPGQR1} where other types of geometries have been found. Finally, column five refers to papers where the group under study leads to a viable model both for quark and lepton mixing patterns.  The recent reference 
\cite{Chaber2018} is taken apart from the other references singled out with the index $\dag$ in the tables. It is based on the alternative concept of a two-Higgs-doublet model.

\subsection{Groups in the series $\Delta(6n^2)$ and more groups}
\label{deltaseries}

An important paper dealing with the series $\Delta(6n^2)\cong \mathbb{Z}_n^2 \rtimes S_3$ as a good model for lepton mixing is \cite{King2013}. A group in this series has to be spontaneously broken into two subgroups, one abelian subgroup $\mathbb{Z}_m^T$ in the charged lepton sector and a Klein subgroup $\mathbb{Z}_2^S \times \mathbb{Z}_2^U$ in the neutrino sector (with neutrinos seen as Majorana particles). The superscripts $S$, $T$ and $U$ refer to the generators of their corresponding $\mathbb{Z}_m$ group in the diagonal charged lepton basis. In this particular model, there is trimaximal lepton mixing with (so called reactor angle) $\theta_{13}$ fixed up to a discrete choice, an oscillation phase zero or $\pi$ and the (so-called atmospheric angle) $\theta_{23}=45^\circ \pm \theta_{13}/\sqrt{2}$.

It is shown in \cite[Table I]{Holthausen2013} that two groups in this series with $n=10$ and $n=16$ provide leading order leptonic mixing patterns within $3$-sigma of current best fit with acceptable entries in the CKM matrix. The small group $(648,259)=D_{18,6}^{(1)}$ also satisfies this requirement. Additionally, if one accepts that neutrinos are Dirac particles then the residual symmetry group of neutrino masses is no longer restricted to the Klein group but may be any abelian group. In such a case, four small groups that are $\Delta(6\times 5^2)$ and small groups $(162,10)$, $(162,12)$ and $(162,14)=D_{9,3}^{(1)}$ predict acceptable entries for the quark and lepton mixing matrices \cite[Table II]{Holthausen2013}.  It is noticeable that our small selection of groups (from requirements (a) and (b) include all of them except for the group $(162,10)$ whose two-dimensional representations are not MICs.

Still assuming that neutrinos are Dirac particles and with loose enough constraints on $V_{us}$, paper \cite{YaoDing2015} include $\Delta$-groups with $n=9$ (it does not lie in our Table \ref{table1}) and $n=14$ in their selection, as well as groups $(648,259)$, $(648,260)$ and $(648, 266)$, the latter groups are in our Table \ref{table2}.
Additional material \cite{YaoDingadd} provides very useful information about the ability of a group to be a good candidate for modeling the mixing patterns. According to this reference, the groups $\Delta(6 \times n^2)$ with $n=10$, $11$, $14$ and $18$, and small groups $(972,64)$ and $(972,245)$, that are in our tables  also match Dirac neutrinos with a $3$-sigma fit and quark mixing patterns for triplet assignment.

Three extra groups $(120,5)$ (the binary icosahedral group $SL(2,5)=2I$), $(360,51)=\mathbb{Z}_3 \times SL(2,5)$ and $(480,221)=SL(2,5).\mathbb{Z}_4$ in our tables, whose signature has a factor equal to the binary icosahedral group $2I$, can be assigned with a doublet and a singlet for quarks but cannot be generated by the residual symmetries in the lepton sector.

\subsection{Exceptional subgroups of $SU(3)$}
\label{exceptionalseries}

The viability of so-called exceptional groups of $SU(3)$ for lepton mixings have been studied in \cite{Hagedorn2014} by assuming neutrinos to be either Dirac or Majorana particles. These subgroups  are listed according to the number of their conjugacy classes in Table \ref {tableexcept}. They are $\Sigma(60)\cong A_5$ (a subgroup of $SO(3)$), $\Sigma(168) \cong PSL(2,7)$, $\Sigma(36 \times 3)$, $\Sigma(72 \times 3)$, $\Sigma(360 \times 3)$ and $\Sigma(216 \times 3)$. Only group $\Sigma(360 \times 3)$ has Klein subgroups and thus supports a model with neutrinos as Majorana particles. Group $\Sigma(216 \times 3)$ is already in our Table \ref{table1} and potentially provides a valid model for quark/lepton mixings by assuming neutrinos are Dirac particles.

According to our Table \ref {tableexcept}, all these exceptional groups have informationally complete characters as regard to most of their faithful three-dimensional representations. Another useful information is about groups $\Sigma(60)$ and $\Sigma(360 \times 3)$ that are informationally complete as regard to their five-dimensional representations.  Models based on the $A_5$ family symmetry are in \cite{Varzielas2014,LiDing2015}.

\small
\begin{table}[h]
\begin{center}
\begin{tabular}{||c|l||c|l|c|l|c|l|c|l|c|l|c|l||}
\hline 
\hline
Group & d&  &  &  & & & & & & &&& \\
\hline
(60,5), $\Sigma(60)$ &5& 1 &{\bf 3} &{\bf 3}  &{\bf 4} &{\bf 5} & & & & &&& \\
5-dit &.& 5 &$d^2$ &$d^2$  &$d^2$ &$d^2$ & & & & &&& \\
\hline
(168,42), $\Sigma(168)$&6 &1&{\bf 3}  & {\bf 3} & {\bf 6} &{\bf 7}&{\bf 8} & & & &&& \\
6-dit &. & 6& $d^2$ & $d^2$ &33  &33 & 33& & & & && \\
\hline
(108,15), $\Sigma(36\times 3)$ & 14& 1 &1  &1  &1 &{\bf 3} & {\bf 3}&{\bf 3} & {\bf 3}&{\bf 3} & {\bf 3}& {\bf 3}&{\bf 3}\\
14-dit & .  &14  &166  &181 &181 &195 &195 &$d^2$ &$d^2$ &$d^2$&$d^2$& $d^2$& $d^2$\\
. & .   &4  &4 & & & & & &&&&& \\
. &.    &154  &154 & & & & & & & & & &  \\
\hline
(216,88), $\Sigma(72\times 3)$ & 16& 1 & 1 & 1 & 1& 2& {\bf 3} &{\bf 3} &{\bf 3} & {\bf 3}&{\bf 3}&{\bf 3}&{\bf 3} \\
16-dit & .&16  &175  &175  &157 & 233&$d^2$ &$d^2$ &$d^2$ &$d^2$ &$d^2$&$d^2$&$d^2$ \\
2Quartits & .&16  &121  &149  &125 & 200&$d^2$ &$d^2$ &$d^2$ &$d^2$ &$d^2$&$d^2$&$d^2$ \\
. & .& {\bf 3} & {\bf 3} &{\bf 3}  & 8& & & & & &&& \\
16-dit & .& $d^2$ & 222 & 222 &144 & & & & & &&& \\
2Quartits & .& $d^2$ &118 & 118 &144 & & & & & &&& \\
\hline
(1080,260), $\Sigma(360\times 3)$& 17& 1 &{\bf 3}  & {\bf 3} &{\bf 3} &{\bf 3} &5 &5 &{\bf 6}&{\bf 6} &8&8& 9\\
 17-dit& .& 17 &$d^2$  & $d^2$ &$d^2$ &$d^2$ &$d^2$ &$d^2$ &$d^2$ &$d^2$ &$d^2$&$d^2$&$d^2$ \\
. & .&{\bf 9}  & {\bf 9} & 10 & {\bf 15}& {\bf 15} & & & & &&& \\
. & .&$d^2$  &$d^2$  & $d^2$ &$d^2$ &$d^2$ & & & & &&& \\
\hline
(648,532),$\Sigma(216\times 3)$& 24& 1 & 1 & 1  &2 &2 &2 &3 &{\bf 3} &{\bf 3} &{\bf 3} &{\bf 3} &{\bf 3} \\
 24-dit & .&24  & 527 &527  &562 & $d^2$&$d^2$ & 560&$d^2$ & $d^2$&$d^2$ & $d^2$&$d^2$ \\
. & .&{\bf 3}&{\bf 6} &{\bf 6}&{\bf 6}  &{\bf 6}  &{\bf 6}  &{\bf 6} & 8& 8& 8& {\bf 9}& {\bf 9}  \\
. & .&$d^2$&$d^2$ &$d^2$& $d^2$ &$d^2$  &$d^2$  &$d^2$ &564 &$d^2$ &$d^2$ &552 &552   \\
\hline
\hline
\end{tabular}
\caption{Exceptional subgroups of $SU(3)$. For each group and each character the table provides the dimension of the representation and the rank of the Gram matrix obtained under the action of the corresponding Pauli group. Bold characters are for faithful representations.} 
\label{tableexcept}
\end{center}
\end{table}
\normalsize

\begin{table}[h]
\begin{center}
\begin{tabular}{|c|l|c|l|c|l|}
\hline 
\hline
Group & Name or signature & cc &  Graph & Ref  \\
\hline
SmallGroup(726,5)   & $\Delta(6 \times 11^2)$ & 38 & $K_{11}^3$ &\cite{Jurciukonis2017,YaoDing2015}\\ 
SmallGroup(648,259) & $(\mathbb{Z}_{18} \times \mathbb{Z}_6)\rtimes S_3$, $D_{18,6}^{(1)}$  & 49 & $K_{18}^3$&\cite{Holthausen2013,Jurciukonis2017,YaoDing2015,LiYaoDing2016}\\ 
SmallGroup(648,260) & $\mathbb{Z}_3^2 \rtimes \mbox{SmallGroup}(72,42)$ & .& .& .\\ 
SmallGroup(648,266) &. & .& $K_6^3$& \cite{Jurciukonis2017}\\ 
SmallGroup(1176,243) &$\Delta(6 \times 14^2)$ & 59& $K_{14}^3$& \cite{Jurciukonis2017,YaoDing2015}\\
SmallGroup(972,64) & $\mathbb{Z}_9^2 \rtimes \mathbb{Z}_{12}$ &62 & $K_{36}^3$ & .\\
SmallGroup(972,245) & $\mathbb{Z}_9^2 \rtimes (\mathbb{Z}_2 \times S_3)$ &. & $K_{18}^3$ & \cite{YaoDing2015}\\
SmallGroup(1536,408544632) &$\Delta(6 \times 16^2)$ & 68& ? &\cite{Holthausen2013,Jurciukonis2017,King2013}\\  
SmallGroup(1944,849) &$\Delta(6 \times 18^2)$ & 85&$K_{18}^3$  &\cite{Jurciukonis2017,YaoDing2015}\\            
\hline
\hline
\end{tabular}
\caption{List of considered groups with number of conjugacy classes $cc > 36$ that satisfy rule (a) (presumably (b) as well) and have been considered before as valid groups for quark/lepton mixing. A reference is given at the last column if a viable model for quark or/and lepton mixings can be obtained. The question mark means that the minimal permutation representation could not be obtained.} 
\label{table2}
\end{center}
\end{table}

\section{Generalized CP symmetry, CP violation}
\label{CPseries}

Currently, many models focus on the introduction of a generalized CP symmetry in the lepton mixing matrix \cite{ChenChuliaDing2018,LiDing2015,Rong2020}. 
The Dirac CP phase $\delta_{CP}=\delta_{13}$ for leptons is believed to be around $-\pi/2$. A set of viable models with discrete symmetries including generalized CP symmetry has been derived in \cite{YaoDing2016}. Most finite groups used for quark/lepton mixings  without taking into account the CP symmetry do survive as carrying generalized CP symmetries. It is found that two extra groups $(294,7)=\Delta(6\times 7^2)$ and $(384,568)=\Delta(6 \times 8^2)$, that have triplet assignments for the quarks, can be added. This confirms the relevance of $\Delta$ models in this context. Group $(294,7)$ was added to our short Table \ref{tablefaithshort} where we see that all of its two- and three-dimensional characters are informationally complete. 

A generalized CP symmetry should not be confused with a \lq physical' CP violation as shown in Reference \cite{Chenetal2014}. A \lq physical' CP violation is a prerequisite for baryogenesis that is the matter-antimatter asymmetry of elementary matter particles. The generalized CP symmetry was introduced as a way of reproducing the absolute values of the entries in the lepton and quark mixing matrices and, at the same time, explaining or predicting the phase angles. A physical CP violation, on the other hand, exchanges particles and antiparticles and its finite group picture had to be clarified.

It is known that the exchange between distinct conjugacy classes of a finite group $G$ is controlled by the outer automorphisms $u$ of the group. Such (non trivial) outer automorphisms have to be class-inverting to correspond to a physical CP violation \cite{Chenetal2014}. This is equivalent to a relation obeyed by the automorphism $u:G \rightarrow G$ that maps every irreducible representation $\rho_{r_i}$ to its conjugate 

$$\rho_{r_i}(u(g))=U_{r_i}\rho_{r_i}(g)^*U_r^\dag,~ \forall g \in G ~\mbox{and}~\forall i,$$
with $U_{r_i}$ a unitary symmetric matrix.

A criterion that ensures that this relation is satisfied is in terms of the so-called twisted Frobenius-Schur indicator over the character $\kappa_{r_i}$

$$FS_u^{(n)}(r_i)=\frac{(\dim r_i)^{(n-1)}}{|G|^n}\sum_{g_i \in G} \kappa_{r_i}(g_1 u(g_1)\cdots g_nu(g_n))=\pm 1,~ \forall i,$$

where $n=\mbox{ord}(u)/2$ if $\mbox{ord}(u)$ is even and $n=\mbox{ord}(u)$ otherwise.

Following this criterion there are three types of groups

1. the groups of type I: there is at least one representation $r_i$ for which $FS_u^{(n)}(r_i)=0$, these groups correspond to a physical CP violation, 

2. groups of type II: for (at least) one automorphism $u \in G$ the $FS_u$'s for all representations are non zero. The automorphism $u$ can be used to define a proper CP transformation in any basis. There are two sub-cases 

Case II A, all $FS_u$'s are $+1$ for one of those u's,

Case II B, some $FS_u$'s are $-1$ for all candidates $u$'s.

A simple program written in the Gap software allows to distinguish these cases \cite[Appendix B]{Chenetal2014}.

Applying this code to our groups in Tables \ref{table1}, \ref{tableexcept} and \ref{table2}, we found that all groups are of type II A or type I.
Type I groups corresponding to a physical CP violation are 

$(216,95)=\Delta(6\times 6^2)$, $(162,44)$, $(216,88)=\Sigma(72 \times 3)$

where we could check that our criteria (a) and (b) apply, the exceptional group $(1080,260)=\Sigma(360,3)$ in Table \ref{tableexcept}  and groups $(972,64)$, $(972,245)$, $(1944,849)=\Delta(6 \times 18^3)$ of Table \ref{table2}.




\section{Conclusion}

Selecting two- and three-dimensional representations of informationally complete characters has been found to be efficient in the context of models of CKM and PMNS mixing matrices. Generalized quantum measurements (in the form of MICs) are customary in the field of quantum information by providing a Bayesian interpretation of quantum theory and leading to an innovative view of universal quantum computing. The aim of this paper has been to see the mixing patterns of matter particles 
with the prism of MICs.
 Our method has been shown to have satisfactorily predictive power for predicting the appropriate symmetries used so far in modeling CKM/PMNS matrices and for investigating the symmetries of $CP$ phases.

 It is admitted that the standard model has to be completed with discrete symmetries or/and to be replaced by more general symmetries such as $SU(5)$ or $E_8\supset SU(5)$, as in F-theory \cite{Meadowcroft}, to account for existing measurements on quarks, leptons and bosons, and the hypothetical dark matter. Imposing the right constraints on the quantum measurements of such particles happens to be a useful operating way.

\section*{Appendix}

\scriptsize
\begin{table}[h]
\begin{center}
\begin{tabular}{||c|l||c|l|c|l|c|l|c|l|c|l|c|l|c|l|c||}
\hline 
\hline
Group & d&  &  &  & & & & & & & & & & & &\\
\hline
(24,12) & 5& 1 & 1 & 2 & {\bf 3} & {\bf 3} & & & & & & &&&& \\
$5$-dit & . & 5 & 21 & $d^2$ & $d^2$ & $d^2$ & & & & & & &&&&\\        
\hline
(120,5) & 9& 1 & {\bf 2} & {\bf 2} & 3 & 3 & 4 & {\bf 4}& 5& {\bf 6}& & & &&&\\
$9$-dit & . & 9 & $d^2$ & $d^2$ & $d^2$ & $d^2$ & $d^2$&$d^2$& 79& $d^2$& & & &&&  \\  
2QT & . & 9 & $d^2$ & $d^2$ & $d^2$ & $d^2$ & $d^2$&$d^2$& $d^2$& $d^2$& & & &&&  \\    
\hline
(150,5) & 13& 1  &1  &2  &{\bf 3} & {\bf 3}& {\bf 3}& {\bf 3}& {\bf 3}& {\bf 3}& {\bf 3}& {\bf 3}&{\bf 6} & {\bf 6}&& \\
13-dit & .& 13  &157  & $d^2$  & $d^2$ &  $d^2$&  $d^2$&  $d^2$&  $d^2$&  $d^2$&  $d^2$&  $d^2$& $d^2$ &  $d^2$&& \\
\hline
(72,42) & 15& 1 &1  &1  & 1& 1& 1& 2&2 &2 &3 & 3&{\bf 3} &{\bf 3}&{\bf 3}&{\bf 3}\\
15-dit & .&15  &203  &209  &209 &195 &195 &219 &$d^2$ & $d^2$&$d^2$ &$d^2$&$d^2$& $d^2$&$d^2$ &$d^2$\\
\hline
(216,95) & 19& 1 & 1 & 2 &2 &2 &2 &3 &3 &3 &3 & {\bf 3}&{\bf 3} &{\bf 3} &{\bf 3} &3\\
19-dit& .& 19 & 343 & 357 &359 &355 &$d^2$  &$d^2$  &$d^2$  &$d^2$  &$d^2$  &$d^2$  &$d^2$  &$d^2$  &$d^2$  &$d^2$ \\
. & .& 3    & 6     & {\bf 6} & {\bf 6}& & & & & & & & & & &\\
. & .&$d^2$ &$d^2$  & $d^2$   & $d^2$  & & & & & & & & & & & \\
\hline
(294,7) &20 & 1  &1  &2  & {\bf 3}&{\bf 3} & {\bf 3}&{\bf 3} &{\bf 3} &{\bf 3} &{\bf 3} &{\bf 3} & {\bf 3}&{\bf 3} & {\bf 3}&{\bf 3}\\
20-dit & .&20  & 349 &388  &$d^2$ &$d^2$ &$d^2$ & $d^2$& $d^2$& $d^2$& $d^2$& $d^2$& $d^2$&$d^2$ &$d^2$ &$d^2$\\
. & .&{\bf 6}  & {\bf 6} & {\bf 6} & {\bf 6}& {\bf 6}& & & & & & & & & &\\
. & .&390  & 390 & 390 & 398& 398& & & & & & & & & &\\
\hline
(72,3) & 21& 1 & 1 & 1 &1 &1 &1 &1 &1 &1&2&2 &2 &{\bf 2} & {\bf 2}&{\bf 2}\\
21-dit & .&21  & 405 &405  &421&421&421 &421 &421 &421 &$d^2$ & $d^2$& $d^2$&$d^2$ & $d^2$&$d^2$\\
. & .& {\bf 2} &{\bf 2}  & {\bf 2}  &3     &3     &3     & & & & & & & & &\\
. & .& $d^2$   &$d^2$    &$d^2$     &$d^2$ & $d^2$&$d^2$ & & & & & & & & &\\
\hline
(162,12) & 22& 1 & 1 & 1 &1 & 1&1 & 2&2 &2 & {\bf 3} &{\bf 3} & {\bf 3}&{\bf 3} &{\bf 3} &{\bf 3}\\
22-dit & .&22  &446  &463  &463 &463 & 463&473 &$d^2$ &$d^2$ &$d^2$ &$d^2$ & $d^2$& $d^2$& $d^2$&$d^2$\\
. & .&{\bf 3} & {\bf 3} & {\bf 3} &{\bf 3} &{\bf 3} & {\bf 3}& 6& & & & & & & &\\
. & .& $d^2$ &$d^2$  &$d^2$  & $d^2$& $d^2$& $d^2$& 198& & & & & & & &\\
\hline
(162,14) & 22&1  & 1 & 1 &1 &1 & 1&2 &2 & 2& {\bf 3} &{\bf 3}&{\bf 3} &{\bf 3} &{\bf 3} &{\bf 3}\\
22-dit & .& 22 & 444 & 461 &463 & 461&463 &473 &$d^2$ &$d^2$ &$d^2$ &$d^2$&$d^2$ &$d^2$ &$d^2$ &$d^2$\\
. & .&{\bf 3}  &{\bf 3}  &{\bf 3} &{\bf 3} &{\bf 3} &{\bf 3}&6 & & & & & & & &\\
. & .&$d^2$  &$d^2$  &$d^2$  &$d^2$&$d^2$ & $d^2$&198 & & & & & & & &\\
\hline
(648,532) & 24& 1 & 1 &1  &2 &2 &2 &3 &{\bf 3} &{\bf 3} &{\bf 3} &{\bf 3} &{\bf 3} & {\bf 3}&{\bf 6} &{\bf 6}\\
24-dit & .&24  & 527 &527  &562 & $d^2$&$d^2$ & 560&$d^2$ & $d^2$&$d^2$ & $d^2$&$d^2$ & $d^2$&$d^2$ &$d^2$\\
3QB-QT & .&24 & 500  & 500 &476  &568 &568 &448 &$d^2$ &$d^2$ & $d^2$&$d^2$ &$d^2$ &$d^2$ & $d^2$&$d^2$ \\
. & .&{\bf 6}  &{\bf 6}  &{\bf 6}  &{\bf 6} & 8& 8& 8& {\bf 9}& {\bf 9}& & & & & &\\
24-dit & .& $d^2$ &$d^2$  &$d^2$  &$d^2$ &564 &$d^2$ &$d^2$ &552 &552 & & & & & &\\
3QB-QT & .&$d^2$ &$d^2$ &$d^2$  &$d^2$  &448 & 560& 560&510 &510 & & & & & & \\
\hline
(648,533) & 24&1  &1  &1  &2 & 2& 2&3 &{\bf 3} &{\bf 3} &{\bf 3} &{\bf 3} &{\bf 3} &{\bf 3} &{\bf 6} &{\bf 6}\\
24-dit & .& 24 &539  &539  &562 &$d^2$ &$d^2$ &514 &$d^2$&$d^2$ &$d^2$ &574 &574 & $d^2$&$d^2$ &$d^2$\\
3QB-QT & .& 24  &532  &532  &481 &572 & 572&452 & 572&568 & 568&570 &570 &572 &575 &$d^2$\\
.& .& {\bf 6} &{\bf 6}  & {\bf 6} & {\bf 6}& 8& 8& 8&{\bf 9} &{\bf 9} & & & & & &\\
24-dit & .& $d^2$ & $d^2$ & $d^2$ &$d^2$ &563 & $d^2$&$d^2$ &478 &478& & & & & &\\
3QB-QT & .&$d^2$  & 573 &573  &575 &488 &560 &560 &520 &520 & & & & & &\\
\hline
\hline
\end{tabular}
\caption{\small Small groups considered in our Table \ref{table1}. For each group and each character the table provides the dimension of the representation and the rank of the Gram matrix obtained under the action of the corresponding Pauli group. Bold characters are for faithful representations. According to our demand, each selected group has both $2$- and $3$-dimensional characters (with at least one of them faithful) that are magic states for an informationally complete POVM (or MIC), with the  rank of Gram matrix equal to $d^2$. The Pauli group performing this action is in general a $d$-dit but is a $2$-qutrit (2QT) for the group $(120,5)=SL(2,5)=2I$, a $3$-qutrit (2QT) for the group $(360,51)=\mathbb{Z}_3 \times SL(2,5)$ or may be a three-qubit/qutrit (3QB-QT) for the groups $(648,532)$ and $(648,533)$.  } 
\label{tablefaith}
\end{center}
\end{table}

\scriptsize
\begin{table}[h]
\begin{center}
\begin{tabular}{||c|l||c|l|c|l|c|l|c|l|c|l|c|l|c|l|c||}
\hline
\hline
Group & d&  &  &  & & & & & & & & & & & &\\
\hline
(120,37) & 25& 1 &1  &1  &1 &1 &1 &1 &1 &1 &1 &2 &2 &2 &2 &2\\
25-dit & .&25  &601  & 601 &601 &601 & 601&601 &601 &601 &601 &623 &$d^2$ &$d^2$ & $d^2$&$d^2$\\
. & .&3  &3  &{\bf 3}  &{\bf 3} &{\bf 3} &{\bf 3} &{\bf 3} &{\bf 3} &{\bf 3}& {\bf 3}& & & &&\\
. & .& $d^2$ &$d^2$  &$d^2$  &$d^2$ &$d^2$ &$d^2$ & $d^2$&$d^2$ &$d^2$ &$d^2$ & & & & &\\
\hline
(360,51) & 27& 1 & 1 & 1 & 2& 2&{\bf 2} & {\bf 2}&{\bf 2} &{\bf 2} &3 &3 &3 & 3& 3&3\\
3QT & .&27  &613  &613  & $d^2$&$d^2$ &$d^2$ &$d^2$ &$d^2$ &$d^2$ & $d^2$& $d^2$& $d^2$& $d^2$& $d^2$&$d^2$\\
. & .& 4 & 4 & 4 &{\bf 4} &4 &{\bf 4} &5 &5 &5 &6 &{\bf 6} &{\bf 6}& & &\\
.& .& 727 & 725 &727  &727 &727 &727 &727 &727 &727 &727 &727 &727 & & &\\
\hline
(162,44) & 30& 1 &1  & 1 &1& 1&1 & 2&2 &2 &2&2 &2& 2& 2&2\\
30-dit & .&31  & 826 & 861 &871 & 861 &871 &883 &877 &879 &883 & 898& $d^2$&$d^2$&$d^2$ &898\\
. & .& 2 &2  &2  &{\bf 3} & {\bf 3}& {\bf 3}&{\bf 3} &{\bf 3} &{\bf 3} & {\bf 3}& {\bf 3}&{\bf 3}&{\bf 3} & {\bf 3}&{\bf 3}\\
. & .& 898 &898  &$d^2$  &$d^2$ &$d^2$ &$d^2$ & $d^2$&$d^2$ &$d^2$ & $d^2$&$d^2$ &$d^2$ &$d^2$ &$d^2$&$d^2$\\
\hline
(600,179) & 33& 1 & 1 & 2 &3 &3 &3 &3 &{\bf 3} &{\bf 3} & {\bf 3}& 3&{\bf 3} &3 &3 &{\bf 3}\\
33-dit & .&33  &1041  & $d^2$ & $d^2$&$d^2$ & $d^2$&$d^2$ &$d^2$ & $d^2$&$d^2$ &$d^2$ &$d^2$ &$d^2$ &$d^2$ &$d^2$\\
. & .&  {\bf 3} &{\bf 3}  &3 &3 &{\bf 3} &3 &{\bf 6} &6 &{\bf 6} &{\bf 6} &{\bf 6} &{\bf 6}& {\bf 6}&{\bf 6}&{\bf 6}\\
. & .&$d^2$  &$d^2$  &$d^2$  &$d^2$ &$d^2$ & $d^2$&$d^2$ &$d^2$ &$d^2$ &$d^2$ &$d^2$ &$d^2$ &$d^2$ &$d^2$ &$d^2$\\
. & .  & {\bf 6} & 6 &{\bf 6} & & & & & & & & & & & &\\
. & .&$d^2$  &$d^2$  &$d^2$  & & & & & & & & & & & &\\
\hline 
(168,45) & 35&1  & 1 &1  &1 &1 &1 &1 &1 &1 &1 &1 &1 & 1&1&2\\
35-dit & .& 35 &1175  &1191  &1191 &1191 &1191 &1191 &1191 &1191 &1191 & 1191& 1191& 1191& 1191&$d^2$\\
. & .&2  &2  &2 &2 &2 &2 &3 &3 & {\bf 3} & {\bf 3}&{\bf 3} &{\bf 3} &{\bf 3} &{\bf 3} &{\bf 3}\\
. & .&$d^2$  &$d^2$  & $d^2$ &$d^2$ &$d^2$&$d^2$ & $d^2$&$d^2$ &$d^2$ &$d^2$ &$d^2$ &$d^2$ &$d^2$ &$d^2$ &$d^2$\\
. & .& {\bf 3} & {\bf 3} &{\bf 3}  &{\bf 3} &{\bf 3} & & & & & & & & & &\\
. & .& $d^2$ & $d^2$ &$d^2$  & $d^2$&$d^2$ & & & & & & & & & &\\
\hline
(480,221) & 36& 1 & 1 &1  & 1& {\bf 2}&{\bf 2} &{\bf 2} & {\bf 2}&{\bf 2} &{\bf 2} &{\bf 2} &{\bf 2} &3 & 3&3\\
36-dit & .& 36 & 36 &1085  &1185 &1184 &$d^2$ &$d^2$ & $d^2$& $d^2$& $d^2$& $d^2$&$d^2$ &1278 &1278 &1278\\
. & .&  3 &3 & 3& 3& 3&4 &4 &4 &4 &{\bf 4} &{\bf 4}&{\bf 4} &{\bf 4}&5 & 5\\
. & .& 1278  &$d^2$  &$d^2$ & $d^2$& $d^2$&1275 & 1278&$d^2$ &$d^2$&$d^2$&$d^2$ &$d^2$ &$d^2$ &1277 &1273\\
. & .& 5  &5 &{\bf 6} &{\bf 6} &{\bf 6} &{\bf 6} & & & & & & & & &\\
. & .&1294   &1294  &1295 &1295 &1295 &1295 & & & & & & & & &\\
\hline
\hline
\end{tabular}
\caption{\small The following up of Table \ref{tablefaith}.  } 
\label{tablefaith2}
\end{center}
\end{table}

\end{document}